\documentclass[11pt]{article}

\usepackage{amsmath}
\usepackage{graphicx}
\usepackage{amsfonts}
\usepackage{amssymb}
\usepackage{epsfig}
\usepackage{color}
\usepackage{psfrag}
\usepackage{epstopdf}

\newcommand{\EQ}{\begin{equation}}
\newcommand{\EN}{\end{equation}}
\newcommand{\be}{\begin{equation}}
\newcommand{\ee}{\end{equation}}
\newcommand{\bea}{\begin{eqnarray}}
\newcommand{\eea}{\end{eqnarray}}

\begin{document} \setcounter{page}{0}
\topmargin 0pt
\oddsidemargin 5mm
\renewcommand{\thefootnote}{\arabic{footnote}}
\newpage
\setcounter{page}{0}
\topmargin 0pt
\oddsidemargin 5mm
\renewcommand{\thefootnote}{\arabic{footnote}}
\newpage
\begin{titlepage}
\begin{flushright}
\end{flushright}
\vspace{0.5cm}
\begin{center}
{\large {\bf Order parameter profiles in presence of topological defect lines}}\\
\vspace{1.8cm}
{\large Gesualdo Delfino}\\
\vspace{0.5cm}
{\em SISSA -- Via Bonomea 265, 34136 Trieste, Italy}\\
{\em INFN sezione di Trieste}\\
\end{center}
\vspace{1.2cm}

\renewcommand{\thefootnote}{\arabic{footnote}}
\setcounter{footnote}{0}

\begin{abstract}
\noindent
We consider the broken phase of the $n$-vector model in $n+1$ dimensions with boundary conditions enforcing the presence of topological defect lines (Ising domain walls, XY vortex lines, and so on), and use field theory to argue an exact expression for the order parameter.
\end{abstract}
\end{titlepage}

\newpage
\noindent
The $n$-vector model is defined by the Hamiltonian
\EQ
{\cal H}=-J\sum_{<i,j>}{\bf s}_i\cdot{\bf s}_j\,,
\label{H}
\EN
where ${\bf s}_i$ is a $n$-component unit vector located at site $i$ of a regular lattice, and the sum is performed over all pairs of nearest neighboring sites. We will refer to the case $J>0$ in $n+1\geq 2$ dimensions, so that for $J$ larger than a critical value $J_c(n)$ the $O(n)$ symmetry characteristic of the Hamiltonian (\ref{H}) is spontaneously broken, i.e. $|\langle{\bf s}_i\rangle|=v>0$. In the following we consider $J>J_c$, close enough to criticality to allow a continuous description in terms of a $(n+1)$-dimensional Euclidean field theory, which is in turn the continuation to imaginary time of a relativistic theory in $n$ space and one time dimensions. Since both the vacuum manifold and the space boundary correspond to the sphere $S^{n-1}$, 
the relativistic theory possesses particle excitations associated to extended field configurations, with different points on the space boundary mapped onto different vacua. These excitations are kinks for $n=1$ (2D Ising model), vortices for $n=2$ (3D XY model), and so on. Their propagation in imaginary time generates topological defect lines for the Euclidean system.

We denote a point in $(n+1)$-dimensional Euclidean space by $({\bf x},y)$, ${\bf x}=(x_1,\ldots,x_n)$, and by ${\bf s}({\bf x},y)$ the order parameter field, namely the continuous version of ${\bf s}_i$; $y$ is imaginary time. We consider the system in the hypercylinder $|{\bf x}|\leq L$, $|y|\leq R/2$, and fix the boundary conditions ${\bf s}({\bf x},y)={\bf x}/|{\bf x}|\equiv\hat{\bf x}$ on the hypersurfaces $|{\bf x}|=L,\,\,|y|<R/2$, and $0<|{\bf x}|\leq L,\,\,y=\pm R/2$. We then take the limit $L\to\infty$ and denote by $\langle\cdots\rangle_{\cal B}$ the statistical averages with these boundary conditions.

The boundary conditions we fixed at $y=\pm R/2$ play the role of final and initial states of the Euclidean time evolution. Denoting these states by $|B(\pm R/2)\rangle$ and the Hamiltonian of the relativistic quantum system by $H$, we have
\EQ
Z_{\cal B}\equiv\langle B(R/2)|B(-R/2)\rangle=\langle B(0)|e^{-RH}|B(0)\rangle\,.
\label{Z}
\EN
If we decompose the boundary states on a basis of asymptotic states of the bulk theory, the large $R$ asymptotics of a quantity like (\ref{Z}) are determined by the asymptotic state with the smallest energy entering the decomposition. Each state in the latter must contain at least one topological particle able to account for the non-trivial boundary conditions we fixed. Calling $\tau$ the lightest of these topological particles and $m$ its mass, we have\footnote{We normalize states by $\langle\tau({\bf p}',\sigma')|\tau({\bf p},\sigma)\rangle=(2\pi)^n\omega\,\delta_{\sigma,\sigma'}\delta({\bf p}-{\bf p}')$.}
\EQ
|B(\pm R/2)\rangle=e^{\pm\frac{R}{2}\omega}\sum_\sigma\int\frac{d{\bf p}}{(2\pi)^n\omega}\,a_\sigma({\bf p})\,|\tau({\bf p},\sigma)\rangle+\ldots\,,
\label{B}
\EN
where ${\bf p}=(p_1,\ldots,p_n)$ is the momentum of the particle, $\omega=\sqrt{{\bf p}^2+m^2}$ its energy, $\sigma$ a spin label\footnote{Particles can carry spin in more than two dimensions ($n>1$ in the present case).}, $a_\sigma({\bf p})$ an amplitude, and the dots stay for states contributing subleading terms to the large $R$ limit of (\ref{Z}); for $n>1$ these additional states may contain also Goldstone bosons. The choice of boundary conditions implies that $\tau$ is located at ${\bf x}=0$ for $y=\pm R/2$. Substitution of (\ref{B}) into (\ref{Z}) gives
\bea
Z_{\cal B}&\sim &\sum_\sigma\int\frac{d{\bf p}}{(2\pi)^n\omega}\,|a_\sigma({\bf p})|^2e^{-\omega R}\sim\sum_\sigma|a_\sigma(0)|^2\int\frac{d{\bf p}}{(2\pi)^nm}\,e^{-(m+\frac{{\bf p}^2}{2m})R}
\nonumber\\
&=&\frac{\sum_\sigma|a_\sigma(0)|^2}{m}\left(\frac{m}{2\pi R}\right)^{n/2}e^{-mR}\,;
\label{Z1}
\eea
here and below the symbol $\sim$ referred to functions of $R$ indicates omission of terms subleading for $mR$ large.

The expectation value of a field $\Phi({\bf x},y)$ with the bundary conditions we have chosen is
\EQ
\langle\Phi({\bf x},y)\rangle_{\cal B}=\frac{1}{Z_{\cal B}}\,\langle B(R/2)|\Phi({\bf x},y)|B(-R/2)\rangle
\label{vPhi}\,.
\EN
We have in particular
\bea
\langle\Phi({\bf x},0)\rangle_{\cal B}&\sim &\frac{1}{Z_{\cal B}}\sum_{\sigma_1,\sigma_2}\int\frac{d{\bf p}_1}{(2\pi)^n\omega_1}\frac{d{\bf p}_2}{(2\pi)^n\omega_2}\,a_{\sigma_1}^*({\bf p}_1)\,\langle\tau({\bf p}_1,\sigma_1)|\Phi(0,0)|\tau({\bf p}_2,\sigma_2)\rangle \,a_{\sigma_2}({\bf p}_2)\,\nonumber\\
&\times& e^{-\frac{R}{2}(\omega_1+\omega_2)+i{\bf x}\cdot({\bf p}_1-{\bf p}_2)}\nonumber\\
&\sim &\left(\frac{2\pi R}{m}\right)^{n/2}\int\frac{d{\bf p}_1d{\bf p}_2}{(2\pi)^{2n}m}\,F_\Phi({\bf p}_1|{\bf p}_2)\,e^{-\frac{R}{4m}({\bf p}_1^2+{\bf p}_2^2)+i{\bf x}\cdot({\bf p}_1-{\bf p}_2)}\,,
\label{vPhi0}
\eea
where
\EQ
F_\Phi({\bf p}_1|{\bf p}_2)\equiv\frac{\sum_{\sigma_1,\sigma_2}a_{\sigma_1}^*(0)a_{\sigma_2}(0)\,\langle\tau({\bf p}_1,\sigma_1)|\Phi(0,0)|\tau({\bf p}_2,\sigma_2)\rangle}{\sum_\sigma|a_\sigma(0)|^2}
\label{ff}
\EN
is the form factor sum whose behavior at small momenta determines the final form of (\ref{vPhi0}). If we consider the field $\varepsilon\sim{\bf s}\cdot{\bf s}$, $F_\varepsilon(0|0)$ is a constant proportional to the square mass of the particle. Integration of (\ref{vPhi0}) then gives
\EQ
\langle\varepsilon({\bf x},0)\rangle_{\cal B}\sim\frac{F_\varepsilon(0|0)}{m}\left(\frac{2m}{\pi R}\right)^{n/2}\,e^{-\frac{2m}{R}{\bf x}^2}\,;
\label{Geps}
\EN
up to the normalization, this Gaussian determines the mass distribution on the hyperplane $y=0$, and then the probability that the particle trajectory (i.e. the defect line) intersects this hyperplane in a volume element $d{\bf x}$ around the point ${\bf x}$.

Turning to the order parameter field, the boundary conditions imply
\EQ
\lim_{|{\bf x}|\to\infty}\langle{\bf s}({\bf x},0)\rangle_{\cal B}=v\,\hat{\bf x}\,,
\label{c1}
\EN
\EQ
\langle{\bf s}(0,0)\rangle_{\cal B}=0\,.
\label{c2}
\EN
We now argue that the low energy limit of $F_{\bf s}({\bf p}_1|{\bf p}_2)$ takes the form
\EQ
F_{\bf s}({\bf p}_1|{\bf p}_2)\sim -i C_nmv\,\frac{{\bf q}}{|{\bf q}|^{n+1}}+D_n\,|{\bf q}|^{\alpha_n}\,{\bf p}\,,\hspace{.5cm}{\bf p}_1,\,{\bf p}_2\to 0\,,
\label{ffs}
\EN
where $C_n$, $D_n$ and $\alpha_n$ are constants, and we introduced ${\bf p}\equiv({\bf p}_1+{\bf p}_2)/2$ and ${\bf q}\equiv({\bf p}_1-{\bf p}_2)/2$. The single non-constant relativistic invariant that can be built out of the two momenta can be written as $({\bf p}_1-{\bf p}_2)^2-(\omega_1-\omega_2)^2$, and reduces to $4{\bf q}^2$ at low energies. Given that the order parameter and the momenta rotate with the same group, the simplest way to represent the symmetry appears that in which $F_{\bf s}({\bf p}_1|{\bf p}_2)$ is a linear combination of ${\bf q}$ and ${\bf p}$. The coefficients of the combination are  functions of the relativistic invariant which in the low energy limit become powers of $|{\bf q}|$. The part of (\ref{ffs}) proportional to ${\bf p}$, when inserted  in (\ref{vPhi0}) and integrated over $d{\bf p}$ contributes zero by parity. Hence, the large $R$ limit of $\langle{\bf s}({\bf x},0)\rangle_{\cal B}$ is determined only by the part proportional to ${\bf q}$. The specific power $|{\bf q}|^{-(n+1)}$ in (\ref{ffs}) is the one consistent with the asymptotic behavior (\ref{c1}). Indeed, substitution of (\ref{ffs}) into (\ref{vPhi0}) gives
\bea
\langle{\bf s}({\bf x},0)\rangle_{\cal B}&\sim & \left(\frac{2\pi R}{m}\right)^{n/2}\frac{-iC_n2^nv}{(2\pi)^{2n}}\int d{\bf p}d{\bf q}\,\frac{\bf q}{|{\bf q}|^{n+1}}\,e^{-\frac{R}{2m}({\bf p}^2+{\bf q}^2)+2i{\bf x}\cdot{\bf q}}
\nonumber\\
&=&-i\frac{C_nv}{\pi^n}\int d{\bf q}\,\frac{\bf q}{|{\bf q}|^{n+1}}\,e^{-{\bf q}^2+2i\sqrt{\frac{2m}R}{\bf x}\cdot{\bf q}}\,,
\label{grad}
\eea
where in the last line we also rescaled ${\bf q}\to\sqrt{\frac{2m}{R}}\,{\bf q}$. Notice that (\ref{grad}) is the gradient of a function of $|{\bf x}|$, and is then proportional to ${\bf x}$; hence we can work in the frame in which ${\bf x}=(|{\bf x}|,0,\ldots,0)$ and write in general
\EQ
\langle{\bf s}({\bf x},0)\rangle_{\cal B}\sim-i\frac{C_nv}{\pi^n}\,\hat{\bf x}\,\int d{\bf q}\,\frac{q_1}{|{\bf q}|^{n+1}}\,e^{-{\bf q}^2+2izq_1}\,,
\label{temp}
\EN
where we introduced
\EQ
z\equiv \sqrt{\frac{2m}{R}}\,|{\bf x}|\,.
\label{rescaled}
\EN
Since the integral in (\ref{temp}) needs to be regularized at small $|{\bf q}|$, we consider instead
\bea
\partial_z\langle{\bf s}({\bf x},0)\rangle_{\cal B}&\sim &\frac{2C_nv}{\pi^n}\,\hat{\bf x}\,\int d\Omega\int_0^\infty dq\,\cos^2\varphi_1\,e^{-q^2+2izq\cos\varphi_1}
\nonumber\\
&=&\frac{C_nv\sqrt{\pi}}{\pi^n}\,\hat{\bf x}\,\int d\Omega\,\cos^2\varphi_1\,e^{-z^2\cos^2\varphi_1}\,,
\nonumber\\
&=&\frac{C_nv\,\pi^{(n+1)/2}}{\pi^n\Gamma\left(1+\frac{n}{2}\right)}\,\,{}_1F_1\left(\frac{3}{2},1+\frac{n}{2};-z^2\right)\,\hat{\bf x}\,,
\eea
where  we introduced spherical coordinates ${\bf q}=q\left(\cos\varphi_1,\sin\varphi_1\cos\varphi_2,\sin\varphi_1\sin\varphi_2\cos\varphi_3,\ldots,\right.$\\
$\left.\sin\varphi_1\ldots\sin\varphi_{n-2}\cos\varphi_{n-1},\sin\varphi_1\ldots\sin\varphi_{n-1}\right)$, with $\varphi_i$ taking values in $(0,\pi)$ for $i=1,\ldots,n-2$, and in $(0,2\pi)$ for $i=n-1$, $d\Omega=d\varphi_1\ldots d\varphi_{n-1}\,\prod_{k=1}^{n-2}\sin^{n-1-k}\varphi_k$, and the integration over $\varphi_1$ leads to the appearance of the confluent hypergeometric function ${}_1F_1(\alpha,\gamma;y)$, which behaves as $y^{-\alpha}$ for $|y|$ large. Integrating back over $z$ we have
\bea
\langle{\bf s}({\bf x},0)\rangle_{\cal B}&\sim & \frac{\Gamma\left(\frac{n+1}{2}\right)}{\Gamma\left(1+\frac{n}{2}\right)}\,v\,\hat{\bf x}\int_0^z dt\,{}_1F_1\left(\frac{3}{2},1+\frac{n}{2};-t^2\right)
\nonumber\\
&=& v\,\frac{\Gamma\left(\frac{n+1}{2}\right)}{\Gamma\left(1+\frac{n}{2}\right)}\,{}_1F_1\left(\frac{1}{2},1+\frac{n}{2};-z^2\right)z\,\hat{\bf x}\,,
\label{result}
\eea
where we used (\ref{c2}) to fix the integration constant, and (\ref{c1}) to obtain 
\EQ
C_n=\pi^{(n-1)/2}\Gamma\left(\frac{n+1}{2}\right)\,.
\EN
For $n=1$ $F_{\bf s}({\bf p}_1|{\bf p}_2)$ must reduce to a Lorentz scalar, namely it must be $D_1=0$ in   (\ref{ffs}). This result for $F_{\bf s}({\bf p}_1|{\bf p}_2)$ is well known from 2D Ising field theory \cite{BKW,review}, where it is usually quoted in the form $-2iv/(\theta_1-\theta_2)$, with rapidities $\theta_i$ parameterizing momenta as $p_i=m\sinh\theta_i$. 

\begin{figure}[t]
\begin{center}
\includegraphics[width=8cm]{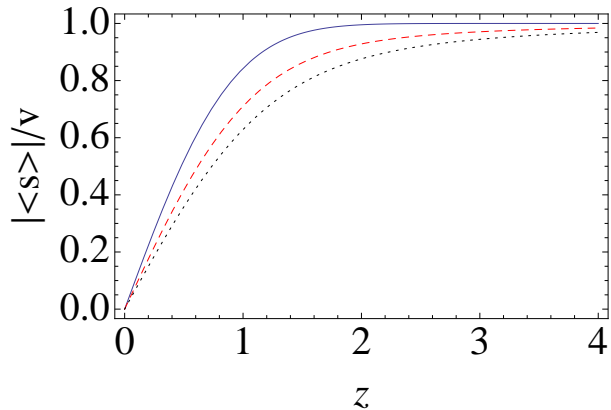}
\caption{Order parameter profiles $|\langle{\bf s}({\bf x},0)\rangle_{\cal B}|/v$ from (\ref{result}) for $n=1$ (2D Ising, continuous curve), $n=2$ (3D XY, dashed curve) and $n=3$ (4D Heisenberg, dotted curve).}
\label{profiles}
\end{center}
\end{figure}

The result (\ref{result}) is illustrated in Fig.~\ref{profiles} for the first few values of $n$. For $n=1$ it reduces to $v\,\mbox{erf(z)}$; this expression, which describes phase separation in the 2D Ising model, was obtained from the exact lattice solution in \cite{Abraham} and from field theory in \cite{DVS}. As for $n>1$, we are not aware of numerical results for comparison with (\ref{result}).

Summarizing, we considered the broken phase of the $(n+1)$-dimensional $n$-vector model with boundary conditions leading to the presence of topological defect lines with fixed endpoints separated by a distance $R$, and argued the exact large $R$ asymptotics of the order parameter on the hyperplane at mid distance between the endpoints. The result has been related to a low energy singularity of the matrix element of the order parameter field on topological particle states. Singularities of this type are known as 'kinematical', in contrast to bound state poles and branch points associated to the opening of scattering channels, and are unwanted in ordinary cases (see \cite{Barton}). This note illustrates their role when the particles correspond to extended configurations of the field.



\end{document}